\documentclass[twocolumn,showpacs,preprintnumbers,amsmath,amssymb]{revtex4}
\usepackage{graphicx}
\usepackage{dcolumn}
\usepackage{bm}

\begin{document}

\preprint{APS/123-QED}

\title{Perturbative analysis of generally nonlocal spatial optical solitons}

\author{Shigen Ouyang}
\author{Qi Guo}
\email{guoq@scnu.edu.cn}
\author{Wei Hu}
\affiliation{Laboratory of Photonic Information Technology, South
China Normal University Guangzhou, 510631, P. R. China}

\date{\today}

\begin{abstract}
In analogy to a perturbed harmonic oscillator, we calculate the
fundamental and some other higher order soliton solutions of the
nonlocal nonlinear Sch\"{o}dinger equation (NNLSE) in the 2nd
approximation in the generally nonlocal case. Comparing with
numerical simulations we show that soliton solutions in the 2nd
approximation can describe the generally nonlocal soliton states
of the NNLSE more exactly than that in the 0th approximation. We
show that for the nonlocal case of an exponential-decay type
nonlocal response the Gaussian-function-like soliton solutions
can't describe the nonlocal soliton states exactly even in the
strongly nonlocal case. The properties of such nonlocal solitons
are investigated. In the strongly nonlocal limit, the soliton's
power and phase constant are both in inverse proportion to the 4th
power of its beam width for the nonlocal case of a Gaussian
function type nonlocal response, and are both in inverse
proportion to the 3th power of its beam width for the nonlocal
case of an exponential-decay type nonlocal response.
\end{abstract}

\pacs{42.65.Tg~,~42.65.Jx~,~42.70.Nq~,~42.70.Df}

\maketitle

\section{\label{sec1}Introduction}
Since Snyder and Mitchell's pioneering work\cite{3}, spatial
solitons propagating in nonlocal nonlinear media have been
investigated experimentally and theoretically in a variety of
configurations and material systems. It is theoretically indicated
that stable spatial bright (dark) soliton states can be admitted
in self-focus(self-defocus) weekly nonlocal media\cite{2} and
Gaussian-function-like bright soliton states can be admitted in
self-focus strongly nonlocal media\cite{3,4}. It has been shown
theoretically that nonlocality drastically modifies the
interaction of dark solitons by inducing a long-range attraction
between them, thereby permitting the formation of stable dark
soliton bound states\cite{11}. The propagation properties of light
beams in the presence of losses in the strongly nonlocal case are
different from that in the local case\cite{25}. By considering the
special case of a logarithmic type of nonlinearity and a Gaussian
function type nonlocal response, the dynamics of beams in
partially nonlocal media\cite{12} and the propagation of
incoherent optical beams\cite{13} are analytically studied. By
using the variational principle, the propagation properties of a
solitary wave in nonlinear nonlocal medium with a power function
type nonlocal response are studied\cite{16}. The modulational
instability of plane waves in nonlocal Kerr media\cite{10,14} and
the stabilizing effect of nonlocality\cite{1} have been studied.
The analogy between parametric interaction in quadratic media and
nonlocal Kerr-type nonlinearities can provide a physically
intuitive theory for quadratic solitons\cite{15}. Some properties
of the strongly nonlocal solitons (SNSs) and their interaction are
greatly different from that in the local case, e. g. two coherent
SNSs with $\pi$ phase difference attract rather than repel each
other\cite{3}, the phase shift of the SNS can be very large
comparing with the local soliton with the same beam width\cite{4},
and the phase shift of a probe beam can be modulated by a pump
beam in the strongly nonlocal case\cite{26}. Employing a Gaussian
ansatz and using a variational approach, the evolution of a
Gaussian beam in the sub-strongly nonlocal case is
studied\cite{27}. Recently it is experimentally shows that
solitons in the nematic liquid crystal(NLC) are SNSs\cite{5,17}.
The team of Assanto has developed a general theory of spatial
solitons in the NLC that exhibiting a nonlinearity with an
arbitrary degree of an effective nonlocality and established an
important link between the SNS and the parametric
soliton\cite{5,17,18,19}. They also experimentally investigated
the role of the nonlocality in transverse modulational
instability(MI) in the NLC\cite{18,19} and observed the optical
multisoliton generation following the onset of spatial
MI\cite{20}. The interaction of SNSs has been experimentally
demonstrated\cite{21,22}, and the possibility of all-optical
switching and logic gating with SNSs in the NLC has been
discussed\cite{23}.

However the theoretical studies on the spatial nonlocal soliton
are mostly focused on the strongly nonlocal
case\cite{3,4,5,17,25,26} and the weekly nonlocal case\cite{2}.
There is a lack of studies on the moderate nonlocal case. On the
other hand, even though a convenient method has been introduced in
references \cite{4,25,26,27} to study the propagation of light
beams in the strongly nonlocal case or even in the sub-strongly
nonlocal case, to employ this method efficiently the nonlocal
response function must be twice differentiable at its center. As
will be shown this method can't deal with the nonlocal case of an
exponential-decay type nonlocal response function that is not
differentiable at its center. In this paper, in analogy to a
perturbed harmonic oscillator, we calculate the fundamental and
some other higher order soliton solutions of the NNLSE in the 2nd
approximation in the generally nonlocal case. Our method presented
here can deal with the nonlocal case of an exponential-decay type
nonlocal response function. Numerical simulations conform that the
soliton solution in the 2nd approximation can describe the
generally nonlocal soliton states of NNLSE more exactly than that
in the 0th approximation. It is shown that for the nonlocal case
of an exponential-decay type nonlocal response the
Gaussian-function-like soliton solutions can't describe the
fundamental soliton states of the NNLSE exactly even in the
strongly nonlocal case, that is greatly different from the case of
a Gaussian function type nonlocal response. The properties of such
nonlocal solitons are investigated. The functional dependence of
such nonlocal soliton's power and phase constant on its beam width
is greatly different from that of the local soliton. Further more
this functional dependence for the nonlocal case of a Gaussian
function type nonlocal response greatly differs from that of an
exponential-decay type nonlocal response. In particular in the
strongly nonlocal limit, the nonlocal soliton's power and phase
constant are both in inverse proportion to the 4th power of its
beam width for the nonlocal case of a Gaussian function type
nonlocal response, and are both in inverse proportion to the 3th
power of its beam width for the nonlocal case of an
exponential-decay type nonlocal response.

\section{\label{first}the fundamental generally nonlocal soliton solution in the 2nd approximation}
Let's consider the (1+1)-D dimensionless nonlocal nonlinear
Sch\"{o}dinger equation(NNLSE)~\cite{2,4,10,13,14,15,16}
\begin{equation}
i{\partial u\over\partial z}+\frac{1}{2}{\partial^2 u\over\partial
x^2}+u\int^{+\infty}_{-\infty}R(x-\xi)|u(\xi,z)|^2d\xi=0,
\label{NNLSE}
\end{equation}
where $u(x,z)$ is the complex amplitude envelop of the light beam,
$x$ and $z$ are transverse and  longitude coordinates
respectively, $R(x)>0$ is the real symmetric nonlocal response
function, and
\begin{equation}
n(x,z)=\int^{+\infty}_{-\infty}R(x-\xi)|u(\xi,z)|^2d\xi
\end{equation}
is the light-induced perturbed refractive index.

As indicated in reference \cite{4}, if $R(x)$ is twice
differentiable at $x=0$ and the 2nd derivative
$R^{\prime\prime}(0)<0$, and if the characteristic nonlocal length
is one order of the magnitude larger than the beam width of the
soliton, the NNLSE~(\ref{NNLSE}) can be simplified to the
following strongly nonlocal model(SNM)
\begin{equation}
i{{\partial u}\over{\partial z}}
+\frac{1}{2}{{\partial^2u}\over{\partial x^2}}
+u\int^{+\infty}_{-\infty}[R_0+\frac{R^{\prime\prime}_0}{2}(x-\xi)^2]|u(\xi,z)|^2d\xi=0,
\label{SNM}
\end{equation}
where $R_0=R(0)$ and $R^{\prime\prime}_0=R^{\prime\prime}(0)$. For
example, for the Gaussian function type nonlocal response function
$R(x)=1/(w_0\sqrt{\pi})\exp\left(-x^2/w_0^2\right)$, when the
characteristic nonlocal length $w_0$ is one order of the magnitude
larger than the beam width, the SNM~(\ref{SNM}) can describe the
NNLSE~(\ref{NNLSE}) very well\cite{4}. However, as will be shown,
when the characteristic nonlocal length and the beam width are in
the same order of the magnitude, the SNM~(\ref{SNM}) can't
describe the NNLSE~(\ref{NNLSE}) very well. The SNM~(\ref{SNM})
can't deal with the generally nonlocal case. Further more, for the
exponential-decay type nonlocal response function
$R(x)=1/(2w_0)\exp(-|x|/w_0)$ which is not differentiable at
$x=0$, we can't get the parameter $R^{\prime\prime}_0$ of the
SNM~(\ref{SNM}). So the SNM~(\ref{SNM}) can't deal with this
nonlocal case of such an exponential-decay type nonlocal response.

The SNM~(\ref{SNM}) allows a Gaussian-function-like bright soliton
solution
\begin{equation}
u_0(x,z)=A\left({{1}\over{\pi\nu^2}}\right)^{1/4}\exp[-{{x^2}\over{2\nu^2}}
-i({{3}\over{4\nu^2}}-R_0A^2 )z] \label{solution-SNM}
\end{equation}
where
\begin{equation}
{{1}\over{\nu^4}}=-R^{\prime\prime}_0A^2,
\end{equation}
and $\nu$ is the beam width of $u_0(x,z)$. The power $P$ and the
phase constant $\gamma$ of $u_0(x,z)$ are given by
\begin{equation}
P=\int^{+\infty}_{-\infty}|u(x,t)|^2dx=A^2,
\end{equation}
\begin{equation}
\gamma=R_0A^2-{{3}\over{4\nu^2}}
\end{equation}
respectively.

In this paper, we define the degree of nonlocality by the ratio of
the characteristic nonlocal length to the beam width of the light
beam and use the phrase \textsl{``generally nonlocal case''} to
refer to the nonlocal case where the degree of nonlocality is
larger than one and less than ten. For the Gaussian function type
nonlocal response function and the soliton
solution~(\ref{solution-SNM}), the degree of nonlocality is
$w_0/\nu$. The larger of $w_0/\nu$, the stronger of the
nonlocality. In fact for a given type of nonlocal response,
soliton solutions with the same degree of nonlocality can be
described in the same way. That can be clarified by taking
transformations\cite{ww}
\begin{equation}
\bar{x}={{x}\over{\kappa}}~~~~~~\bar{z}={{z}\over{\kappa^2}}~~~~~~\bar{u}=\kappa
u~~~~~~~~\bar{R}=\kappa R~.
\end{equation}
Under these transformations, the form of NNLSE~(\ref{NNLSE}) keeps
invariant and the degree of nonlocality keeps invariant too. If we
set $\kappa$ equal to the characteristic nonlocal length of
$R(x)$, the characteristic nonlocal length of $\bar{R}(\bar{x})$
will be scaled to unity and the degree of nonlocality will be
determined only by the beam width of $\bar{u}(\bar{x},\bar{z})$.
In this case the less of the beam width of
$\bar{u}(\bar{x},\bar{z})$, the stronger of the nonlocality. On
the other hand we may also set $\kappa$ equal to the beam width of
$u(x,z)$. If we do this, the degree of nonlocality will be
determined only by the characteristic nonlocal length of
$\bar{R}(\bar{x})$. The larger of the characteristic nonlocal
length of $\bar{R}(\bar{x})$, the stronger of the nonlocality. In
this paper, the characteristic nonlocal length of $R(x)$ and the
beam width of $u(x,z)$ are not scaled to unity.

For the soliton state $u(x,z)$, we have $|u(-x,z)|^2=|u(x,z)|^2$
and $|u(x,z)|=|u(x,0)|$. So for the soliton state $u(x,z)$, by
defining
\begin{equation}
V(x)=-\int^{+\infty}_{-\infty}R(x-\xi)|u(\xi,z)|^2d\xi,
\label{defineV}
\end{equation}
the NNLSE~(\ref{NNLSE}) reduces to
\begin{equation}
i{\partial u\over\partial z}+\frac{1}{2}{\partial^2 u\over\partial
x^2}-V(x)u=0.\label{SE}
\end{equation}
Taking the Taylor's expansion of $V(x)$ at $x=0$, we obtain
\begin{equation}
V(x)=V_0+{{1}\over{2\mu^4}}x^2+\alpha x^4+\beta x^6+\cdots,
\label{V}
\end{equation}
where
\begin{subequations}
\begin{eqnarray}
&V_0=V(0),\\
&{{1}\over{\mu^4}}=V^{(2)}(0),\\
&\alpha={{1}\over{4!}}V^{(4)}(0),\\
&\beta={{1}\over{6!}}V^{(6)}(0).
\end{eqnarray}
\label{parameters}
\end{subequations}
As will be shown, in the generally nonlocal case and the strongly
nonlocal case the parameter $\mu$ can be viewed as the beam width
of the soliton, and when $x<\mu$, the terms $\alpha x^4$ and
$\beta x^6$ are one and two order of the magnitude smaller than
the term $x^2/(2\mu^4)$ respectively. That indicates the effects
of $\alpha x^4$ and $\beta x^6$ on the soliton are considerably
small comparing with the effect of $x^2/(2\mu^4)$ in the generally
nonlocal case. Further more in the generally nonlocal case, the
effects of the $x^8$ power term and the other higher power terms
of the Taylor's series of $V(x)$ on the soliton are far smaller
than the effects of these three lower power terms. For convenience
sake we will neglect such higher power terms in the following
discussions and simply adopt
\begin{equation}
V(x)=V_0+{{1}\over{2\mu^4}}x^2+\alpha x^4+\beta x^6. \label{PV}
\end{equation}
However, as the degree of nonlocality decreases the effects of
$\alpha x^4, \beta x^6$ and other higher power terms become larger
and larger, and when the characteristic nonlocal length is
comparable with or less than the beam width of the soliton, the
$x^8$ power term and other higher power terms are no longer
negligible. For such cases we must take the higher power terms of
the Taylor's series of $V(x)$ into account.

In the generally nonlocal case, substitution of Eq.~(\ref{PV})
into Eq.~(\ref{SE}) yields
\begin{equation}
i{{\partial u}\over{\partial z}}=\left[
-\frac{1}{2}{{\partial^2}\over{\partial
x^2}}+V_0+{{1}\over{2\mu^4}}x^2+\alpha x^4+\beta
x^6\right]u.\label{stationary}
\end{equation}
Taking a transformation
\begin{equation}
u(x,z)=\psi_n(x)\exp[-i(\varepsilon_n+V_0)z],\label{transformation}
\end{equation}
we arrive at
\begin{equation}
\left[-\frac{1}{2}{{d^2}\over{d x^2}}+{{1}\over{2\mu^4}}x^2+\alpha
x^4+\beta x^6\right]\psi_n=\varepsilon_n\psi_n, \label{PSE}
\end{equation}
where the index $n=0,1,2,\cdots$ is the order of the soliton
solution, in particular $n=0$ corresponding to the fundamental
soliton solution and $n=1$ corresponding to the second order
soliton solution and so on. Even though Eq.~(\ref{PSE}) takes the
form of the stationary Schr\"{o}dinger equation, the parameters
$\mu, \alpha ,\beta$ are depended on the soliton solution
$\psi_n(x)$.

If $\alpha=0$ and $\beta=0$, equation~(\ref{PSE}) reduces to the
well-known stationary Schr\"{o}dinger equation for a harmonic
oscillator. Since in the generally nonlocal case the effects of
the terms $\alpha x^4$ and $\beta x^6$ on the soliton are far
smaller than that of the term $x^4/(2\mu^4)$, we view the terms
$\alpha x^4$ and $\beta x^6$ as perturbations in the process of
solving Eq.~(\ref{PSE}). Following the perturbation method
presented in any a text book about quantum mechanics(for example,
seeing \cite{24}), we get for the fundamental soliton solution in
the 2nd approximation
\begin{eqnarray}
&\psi_0(A,\alpha,\beta,x)\approx A({{1}\over{\pi\mu^2}})^{1/4}\exp(-{{x^2}\over{2\mu^2}})~~~~~~~~~~~~~~~~~\nonumber\\
&\times[1+\alpha({{9\mu^6}\over{16}}-{{3\mu^4}\over{4}}x^2-{{\mu^2}\over{4}}x^4)~~~~~~~~~~~~~~~~~~~~~~~~~~~~\nonumber\\
&+\alpha^2(-{{1247\mu^{12}}\over{512}}+{{141\mu^{10}}\over{64}}x^2+{{53\mu^8}\over{64}}x^4
+{{13\mu^6}\over{48}}x^6+{{\mu^4}\over{32}}x^8)\nonumber\\
&+\beta({{55\mu^8}\over{32}}-{{15\mu^6}\over{8}}x^2-{{5\mu^4}\over{8}}x^4-{{\mu^2}\over{6}}x^6)],~~~~~~~~~~~~~~~~~~~~~
\label{psi}
\end{eqnarray}
and
\begin{equation}
\varepsilon_0\approx
{{1}\over{2\mu^2}}+{{3\mu^4\alpha}\over{4}}-{{21\mu^{10}\alpha^2}\over{8}}+{{15\mu^6\beta}\over{8}}.\label{E0}
\end{equation}

In Eqs.~(\ref{psi}) and (\ref{E0}), if we neglect the $\alpha,
\alpha^2, \beta$ terms or neglect the $\alpha^2, \beta$ terms
only, we will get for the fundamental soliton solution in the 0th
approximation or in the 1st approximation respectively.

Substituting Eq.~(\ref{psi}) into Eq.~(\ref{defineV}), we have
\begin{equation}
V(A,\alpha,\beta,x)=-\int^{+\infty}_{-\infty}R(x-\xi)\psi_0^2(A,\alpha,\beta,\xi)d\xi.\label{V-function}
\end{equation}
Keeping in mind Eqs.~(\ref{parameters}), we obtain
\begin{subequations}
\begin{eqnarray}
&{{1}\over{\mu^4}}=V^{(2)}(A,\alpha,\beta,0),\\
&\alpha={{1}\over{4!}}V^{(4)}(A,\alpha,\beta,0),\\
&\beta={{1}\over{6!}}V^{(6)}(A,\alpha,\beta,0).
\end{eqnarray}\label{p-function}
\end{subequations}
For a fixed value of parameter $\mu$, the parameters $A, \alpha,
\beta$ can be found by solving Eqs.~(\ref{p-function}). In the
section Appendix~(\ref{A}), we present a fixed-point method to
numerically calculate these parameters based on
Eqs.~(\ref{p-function}). Hereinabove we have formally presented
the main formulas to calculate the perturbed fundamental generally
nonlocal soliton solution in the 2nd approximation.

\subsection{the nonlocal case of a Gaussian function type nonlocal
response}

As an example, let us consider the nonlocal case of a Gaussian
function type nonlocal response\cite{4,12,13,10,1}
\begin{equation}
R(x)={{1}\over{w_0\sqrt{\pi}}}\exp\left(-{{x^2}\over{w_0^2}}\right).\label{Gauss}
\end{equation}
For the SNM~(\ref{SNM}) and the soliton solution
(\ref{solution-SNM}), we can find the fundamental soliton solution
for such a Gaussian function type nonlocal response in the
strongly nonlocal case
\begin{equation}
u_0(x,z)=({{\sqrt{\pi}w_0^3}\over{2\nu^4}})^{{{1}\over{2}}}({{1}\over{\pi\nu^2}})^{{{1}\over{4}}}e^{-{{x^2}\over{2\nu^2}}-i({{3}\over{4\nu^2}}-{{w_0^2}\over{2\nu^4}})z}.\label{solution-SNM-Gaussian}
\end{equation}
This soliton solution can describe the soliton state of the
NNLSE~(\ref{NNLSE}) exactly in the strongly nonlocal case when the
degree of nonlocality $w_0/\nu>10$, but can't describe the soliton
state in the generally nonlocal case when $w_0/\nu\sim2$.

\begin{figure}
\centering
\includegraphics[totalheight=1.2in]{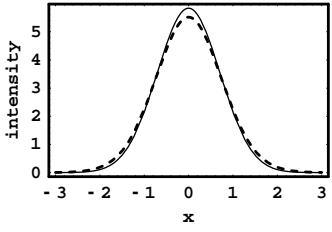}
\caption{\label{fig5}The comparison between
$|\psi_0(A,\alpha,\beta,x)|^2$ (dashing line) and
$|\psi_0(A,0,0,x)|^2$ (solid line). Here $w_0=2, \mu=1, A=3.217,
\alpha=-0.0487, \beta=0.00317$ and the degree of nonlocality
$w_0/\mu=2$.}
\end{figure}

In the generally nonlocal case, the fundamental soliton solution
in the 2nd approximation is described by
$\psi_0(A,\alpha,\beta,x)$ in Eq.~(\ref{psi}). As shown in
Fig.~(\ref{fig5}) when $w_0=2$, $\mu=1$ and $A=3.22$,
$\alpha=-0.0487$, $\beta=0.00317$ numerically calculated by the
fixed-point method presented in the section Appendix~(\ref{A}),
the difference between the fundamental soliton solution in the 2nd
approximation $\psi_0(A,\alpha,\beta,x)$ and that in the 0th
approximation $\psi_0(A,0,0,x)$ is comparatively small. As a
Gaussian function, the power and the beam width of
$\psi_0(A,0,0,x)$ are given by $A^2$ and $\mu$ respectively.
Therefore the power and the beam width of
$\psi_0(A,\alpha,\beta,x)$ are approximatively given by $A^2$ and
$\mu$ respectively too. So in the generally nonlocal case we can
approximately determine the degree of nonlocality by $w_0/\mu$,
and approximately obtain
\begin{eqnarray}
&V(x)\approx-\int^{+\infty}_{-\infty}{{1}\over{w_0\sqrt{\pi}}}\exp\left[-{{(x-\xi)^2}\over{w_0^2}}\right]\psi_0^2(A,0,0,\xi)d\xi\nonumber\\
&=-{{A^2}\over{\sqrt{\pi(\mu^2+w_0^2)}}}\exp\left(-{{
x^2}\over{\mu^2+w_0^2}}\right),~~~~~~~~~~~~
\end{eqnarray}
and
\begin{subequations}
\begin{eqnarray}
&A^2\approx{{\sqrt{\pi}(1+w_0^2/\mu^2)^{3/2}}\over{2\mu}},\\
&V_0\approx-{{(1+w_0^2/\mu^2)}\over{2\mu^2}},\\
&\alpha\approx-{{1}\over{4\mu^6(1+w_0^2/\mu^2)}},\\
&\beta\approx{{1}\over{12\mu^8(1+w_0^2/\mu^2)^2}}.
\end{eqnarray}\label{proxi-p}
\end{subequations}
Using Eqs.~(\ref{proxi-p}) for $w_0=2$ and $\mu=1$, we can find
$A\approx3.14$, $\alpha\approx-{{1}\over{20}}$ and
$\beta\approx{{1}\over{300}}$ that are very close to the
numerically calculated values $A=3.22$, $\alpha=-0.0487$ and
$\beta=0.00317$, and we can find $|{{\alpha
x^4}\over{x^2/(2\mu^4)}}|\approx|{{x^2}\over{10\mu^2}}|<0.1$ and
$|{{\beta
x^6}\over{x^2/(2\mu^4)}}|\approx|{{x^4}\over{150\mu^4}}|<0.007$
for $x<\mu$ that are consistent with the perturbation postulate.

In the strongly nonlocal limit the degree of nonlocality
$w_0/\mu\gg1$, we have
\begin{subequations}
\begin{eqnarray}
&A^2\approx{{\sqrt{\pi}w_0^3}\over{2\mu^4}},\\
&V_0\approx-{{w_0^2}\over{2\mu^4}},\\
&\alpha\approx-{{1}\over{4\mu^4w_0^2}},\\
&\beta\approx{{1}\over{12\mu^4w_0^4}}.
\end{eqnarray}\label{p-Gauss-strongly}
\end{subequations}
As the degree of nonlocality $w_0/\mu$ approaches infinity, the
parameters $\alpha, \beta$ both approach zero, and
$\psi_0(A,\alpha,\beta,x)$ approaches $\psi_0(A,0,0,x)$. In such a
case a Gaussian-function-like strongly nonlocal soliton solution
is obtained.

\begin{figure}
\centering
\includegraphics[totalheight=5in]{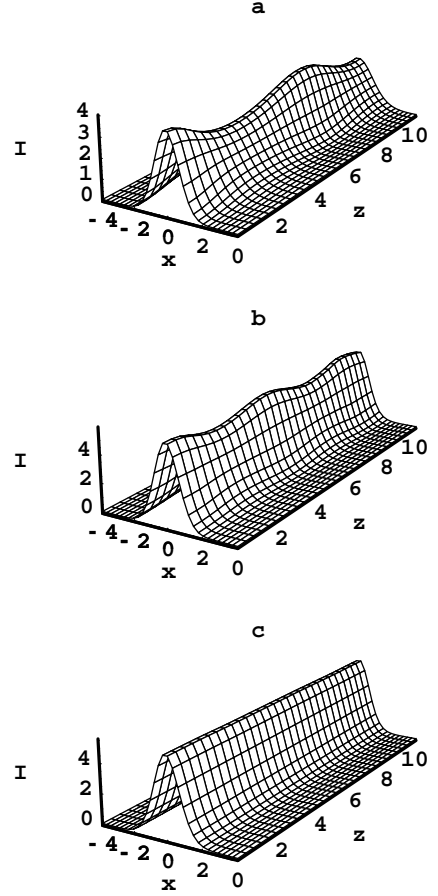}
\caption{\label{fig2}The propagations of light beams in nonlocal
media with a Gaussian function type nonlocal response with
different input intensity profiles that are described by (a)
$|u_0(x,0)|^2$ in Eq.~(\ref{solution-SNM-Gaussian}), (b)
$|\psi_0(A,0,0,x)|^2$ and (c) $|\psi_0(A,\alpha,\beta,x)|^2$
respectively. Here $w_0=2, \mu=1, \nu=1, A=3.217, \alpha=-0.0487,
\beta=0.00317$ and the degree of nonlocality $w_0/\mu=2$.}
\end{figure}

\begin{figure}
\centering
\includegraphics[totalheight=3in]{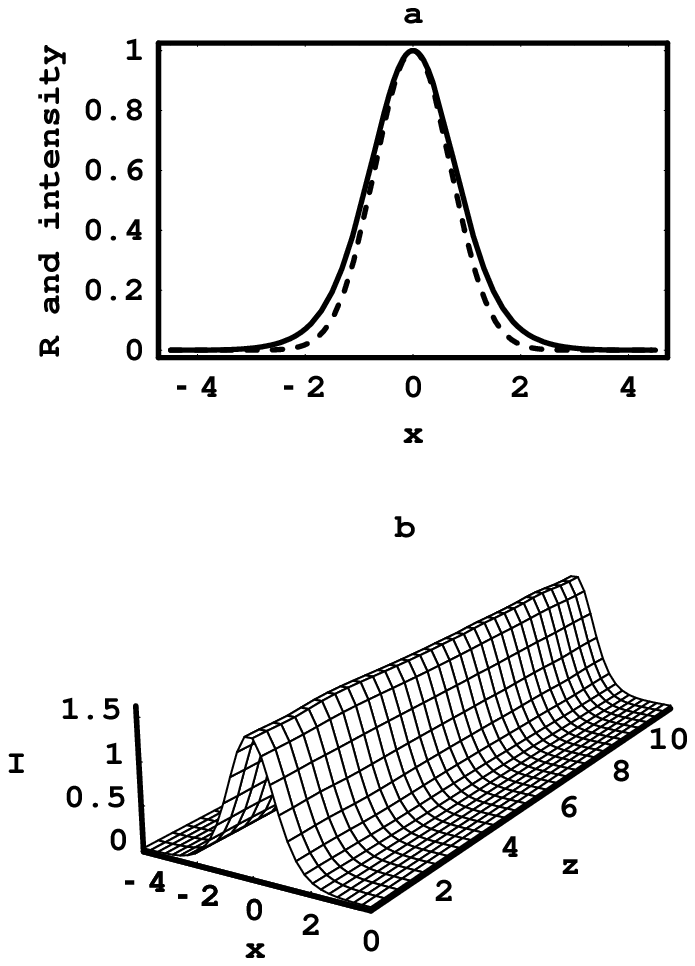}
\caption{\label{fig3}(a) The comparison between $R(x)$ and
$|\psi_0(A,\alpha,\beta,x)|^2$. Dashing line: $R(x)/R(0)$; solid
line: $|\psi_0(A,\alpha,\beta,x)|^2/|\psi_0(A,\alpha,\beta,0)|^2$;
(b)The propagation of the light beam with an input intensity
profile described by $|\psi_0(A,\alpha,\beta,x)|^2$. Here $w_0=1,
\mu=1, A=1.777, \alpha=-0.113, \beta=0.0177$ and the degree of
nonlocality $w_0/\mu=1$.}
\end{figure}

\begin{figure}
\centering
\includegraphics[totalheight=3in]{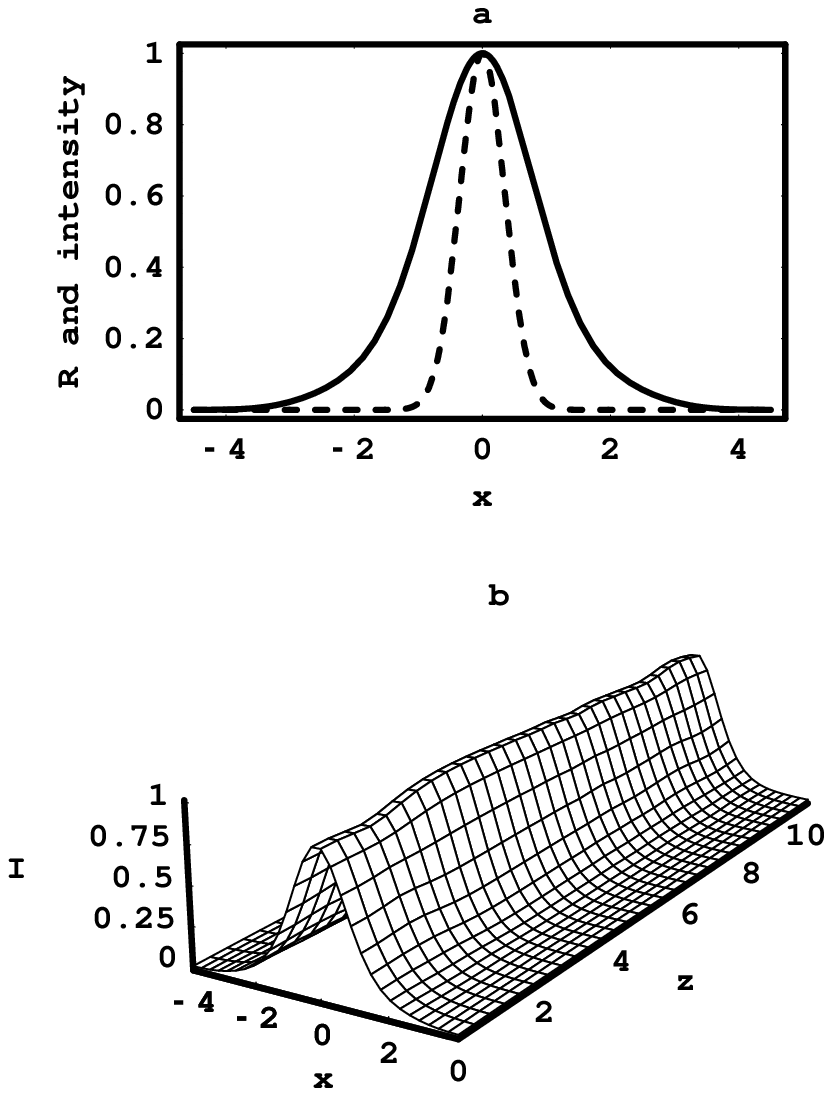}
\caption{\label{fig4}(a) The comparison between $R(x)$ and
$|\psi_0(A,\alpha,\beta,x)|^2$. Dashing line: $R(x)/R(0)$; solid
line: $|\psi_0(A,\alpha,\beta,x)|^2/|\psi_0(A,\alpha,\beta,0)|^2$;
(b)The propagation of the light beam with an input intensity
profile described by $|\psi_0(A,\alpha,\beta,x)|^2$. Here
$w_0=0.5, \mu=1, A=1.403, \alpha=-0.178, \beta=0.0476$ and the
degree of nonlocality $w_0/\mu=0.5$}
\end{figure}

Using the NNLSE~(\ref{NNLSE}) as the evolution equation and using
the numerical simulation method we investigate the propagation of
light beams in nonlocal media with a Gaussian function type
nonlocal response. The numerical simulation method is the
split-step Fourier Method(SSFM)\cite{qq}, the step-size $\Delta
z=0.01$, transversal sampling range $-10\leq x\leq10$ and the
sampling interval $\Delta x=0.1$. With different input amplitude
envelops (the initial data of numerical simulations) that are
described by $u_0(x,0)$ in Eq.~(\ref{solution-SNM-Gaussian}),
$\psi_0(A,0,0,x)$ and $\psi_0(A,\alpha,\beta,x)$ respectively, we
show the propagations of these light beams in Fig.~(\ref{fig2}).
It is indicated that in the generally nonlocal case when the
degree of nonlocality $w_0/\mu=2$, $\psi_0(A,\alpha,\beta,x)$ can
describe the soliton state of the NNLSE~(\ref{NNLSE}) more exactly
than $\psi_0(A,0,0,x)$ and $u_0(x,0)$ in
Eq.~(\ref{solution-SNM-Gaussian}). The soliton solution in the 2nd
approximation $\psi_0(A,\alpha,\beta,x)$ also can describe the
soliton state of the NNLSE~(\ref{NNLSE}) exactly when $w_0/\mu=1$,
that is shown in Fig.~(\ref{fig3}). However when $w_0/\mu=0.5$, as
indicated in Fig.~(\ref{fig4}), $\psi_0(A,\alpha,\beta,x)$ can't
describe the soliton state of the NNLSE~(\ref{NNLSE}) exactly. In
such a case, we must take the higher power terms of the Taylor's
series of $V(x)$ into account and calculate the higher order
approximation. To show how exactly $\psi_0(A,\alpha,\beta,x)$
describe the fundamental soliton state, we define
\begin{subequations}
\begin{eqnarray}
&\theta(z)=\sqrt{{{\int^{+\infty}_{-\infty}|u(x,z)e^{-i\phi(z)}-u(x,0)|^2dx}\over{\int^{+\infty}_{-\infty}|u(x,0)|^2dx}}},\\
&\bar{\theta}={{\int^{l}_{0}\theta(z)dz}\over{l}}.
\end{eqnarray}\label{error}
\end{subequations}
where $e^{i\phi(z)}$ is the phase factor of $u(x,z)$ and for the
fundamental soliton $e^{i\phi(z)}={{u(0,z)}\over{|u(0,z)|}}$. For
a fixed value of $l$, the less of $\bar{\theta}$, the more exactly
$\psi_0(A,\alpha,\beta,x)$ describe the fundamental soliton state.
As shown in table~(\ref{table}), $\psi_0(A,\alpha,\beta,x)$ can
describe the fundamental soltion states exactly when $w_0/\mu>1$.
\begin{table}
\caption{\label{table}Using the numerical simulation method we
calculate $\bar{\theta}$ in Eqs.~(\ref{error}) for the nonlocal
case of the Gaussian function type nonlocal response and for the
nonlocal case of the exponential-decay type nonlocal response.}
\begin{ruledtabular}
\begin{tabular}{ccccccccc}
&&\multicolumn{5}{c}{$\bar{\theta}$}\\
\hline
$\psi_0$\footnotemark[1]& 0.020\footnotemark[3] & 0.0033\footnotemark[4] & 0.0027\footnotemark[5] & 0.0027\footnotemark[6] & 0.0026\footnotemark[7] & 0.0026\footnotemark[8] & 0.0026\footnotemark[9] \\
$\psi_1$\footnotemark[1]& 0.044\footnotemark[3] & 0.013\footnotemark[4] & 0.013\footnotemark[5] & 0.013\footnotemark[6] & 0.013\footnotemark[7] & 0.013\footnotemark[8] & 0.013\footnotemark[9] \\
\hline
$\psi_0$\footnotemark[2]& 0.017\footnotemark[3] & 0.0095\footnotemark[4] & 0.0078\footnotemark[5] & 0.0072\footnotemark[6] & 0.0071\footnotemark[7] & 0.0069\footnotemark[8] & 0.0068\footnotemark[9] \\
$\psi_1$\footnotemark[2]& 0.072\footnotemark[3] & 0.044\footnotemark[4] & 0.037\footnotemark[5] & 0.034\footnotemark[6] & 0.032\footnotemark[7] & 0.031\footnotemark[8] & 0.031\footnotemark[9] \\
$\psi_2$\footnotemark[2]& 0.36\footnotemark[3] & 0.20\footnotemark[4] & 0.17\footnotemark[5] & 0.15\footnotemark[6] & 0.14\footnotemark[7] & 0.14\footnotemark[8] & 0.13\footnotemark[9] \\
\end{tabular}
\end{ruledtabular}
\footnotetext[1]{the nonlocal case of the Gaussian function type
nonlocal response} \footnotetext[2]{the nonlocal case of the
exponential-decay type nonlocal} \footnotetext[3]{the degree of
nonlocality equal to 1} \footnotetext[4]{the degree of nonlocality
equal to 2} \footnotetext[5]{the degree of nonlocality equal to 3}
\footnotetext[6]{the degree of nonlocality equal to 4}
\footnotetext[7]{the degree of nonlocality equal to 5}
\footnotetext[6]{the degree of nonlocality equal to 6}
\footnotetext[9]{the degree of nonlocality equal to 7}
\end{table}

\begin{figure}
\centering
\includegraphics[totalheight=3in]{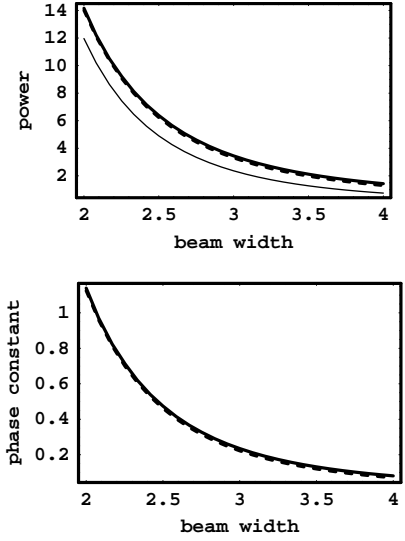}
\caption{\label{fig9}The dependence of the power $P$ and the phase
constant $\gamma$ on the beam width $\mu$. Dashing lines are
described by Eq.~(\ref{Gauss-P}) and Eq.~(\ref{strong-gamma})
respectively; Thin solid line is described by
Eq.~(\ref{strong-P}); Thick solid lines are directly calculated
with parameters $A,\alpha,\beta$ numerically found by the
fixed-point method presented in the section Appendix~(\ref{A}).
Here $w_0=6$.}
\end{figure}

Now let us consider the properties of $\psi_0(A,\alpha,\beta,x)$.
As have been shown, the beam width of $\psi_0(A,\alpha,\beta,x)$
is approximatively given by $\mu$, and its power and phase
constant are approximatively given by
\begin{equation}
P\approx
A^2\approx{{\sqrt{\pi}(1+w_0^2/\mu^2)^{3/2}}\over{2\mu}},\label{Gauss-P}
\end{equation}
\begin{eqnarray}
&\gamma=-V_0-\varepsilon_0~~~~~~~~~~~~~~~~~~~~~~~~~~~~~~~~~~~~~~~~\nonumber\\
&\approx{{1}\over{2\mu^2}}\left[{{w_0^2}\over{\mu^2}}+{{3}\over{8(1+w_0^2/\mu^2)}}+{{1}\over{64(1+w_0^2/\mu^2)^2}}\right]
\end{eqnarray}
respectively. In the strongly nonlocal limit the degree of
nonlocality $w_0/\mu\gg1$, we have
\begin{equation}
P\approx{{\sqrt{\pi}w_0^3}\over{2\mu^4}},\label{strong-P}
\end{equation}
\begin{equation}
\gamma\approx{{w_0^2}\over{2\mu^4}}.\label{strong-gamma}
\end{equation}
That means for a given value of characteristic nonlocal length in
the strongly nonlocal case the power and the phase constant of the
nonlocal soliton are both in inverse proportion to the 4th power
of its beam width. The dependence of the power $P$ and the phase
constant $\gamma$ on the beam width $\mu$ are shown in
Fig.~(\ref{fig9}) for a given value of characteristic nonlocal
length. It is indicated that Eq.~(\ref{Gauss-P}) and
Eq.~(\ref{strong-gamma}) can describe these dependence exactly in
the generally nonlocal case.

To make a comparison with the local soliton, let us consider the
following local nonlinear Schr\"{o}dinger equation(NLSE)\cite{a}
\begin{equation}
i{{\partial u}\over{\partial
z}}+{{1}\over{2}}{{\partial^2u}\over{\partial
x^2}}+|u|^2u=0.\label{NLSE}
\end{equation}
When the characteristic nonlocal length $w_0$ approaches zero, the
Gaussian function type nonlocal response function $R(x)$
approaches the $\delta(x)$ function, and the NNLSE~(\ref{NNLSE})
approaches the NLSE~(\ref{NLSE}). The fundamental soliton of the
NLSE~(\ref{NLSE}) is given by\cite{a}
\begin{equation}
u(x,z)={{1}\over{\eta}}sech({{x}\over{\eta}})\exp\left(i{{z}\over{2\eta^2}}\right),\label{s-NLSE}
\end{equation}
where $\eta$ can be viewed as the beam width of the local soliton.
The power and the phase constant of such a local soliton are given
by
\begin{equation}
P=\int^{+\infty}_{-\infty}|u(x,t)|^2dx={{2}\over{\eta}},
\end{equation}
\begin{equation}
\gamma={{1}\over{2\eta^2}}
\end{equation}
respectively. We can find the power and the phase constant of the
local soliton are in inverse proportion to the 1st and the 2nd
power of its beam width respectively. The functional dependence of
the power and the phase constant of the nonlocal soliton on its
beam width greatly differs from that of the local soltion.

\subsection{the nonlocal case of an exponential-decay type nonlocal
response}

As another example, we investigate the nonlocal case where the
light-induced perturbed refractive index $n(x,z)$ is governed
by\cite{11,10,17}
\begin{equation}
n(x,z)-w_0^2{{\partial^2n(x,z)}\over{\partial
x^2}}-|u(x,z)|^2=0.\label{induced-n}
\end{equation}
It is found that several nonlocal media, for example the nematic
liquid crystal\cite{5,17}, their light-induced perturbed
refractive index can be described by Eq.~(\ref{induced-n}). If the
size of the nonlocal media is much larger than the beam width of
the soliton and the characteristic nonlocal length, the effect of
the boundary condition on the soliton can be negligible and we can
simply assume the size of the nonlocal media is infinity large.
For such a case, equation~(\ref{induced-n}) leads
\begin{equation}
n(x,z)={{1}\over{2w_0}}\int^{+\infty}_{-\infty}\exp\left(-{{|x-\xi|}\over{w_0}}\right)|u(\xi,z)|^2d\xi,
\end{equation}
and we get the exponential-decay type nonlocal
response\cite{1,13,10,11,14}
\begin{equation}
R(x)={{1}\over{2w_0}}\exp\left(-{{|x|}\over{w_0}}\right).\label{exp}
\end{equation}
Since the exponential-decay type nonlocal response function $R(x)$
is not differentiable at $x=0$, the SNM~(\ref{SNM}) can't deal
with this nonlocal case. So we have to use
$\psi_0(A,\alpha,\beta,x)$ to describe the soliton state of the
NNLSE~(\ref{NNLSE}).

\begin{figure}
\centering
\includegraphics[totalheight=3in]{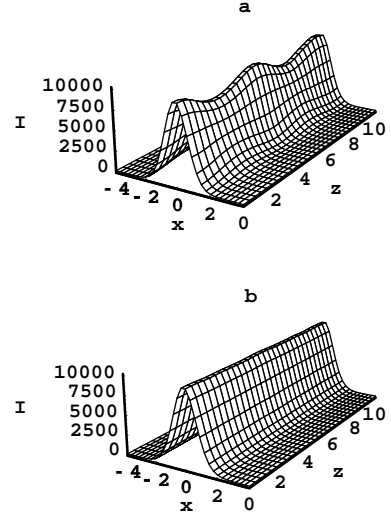}
\caption{\label{fig6}The propagations of light beams with
different input intensity profiles that are described by (a)
$|\psi_0(A,0,0,x)|^2$ and (b) $|\psi_0(A,\alpha,\beta,x)|^2$
respectively. Here $w_0=100, \mu=1, A=138.159, \alpha=-0.0767,
\beta=0.0150$ and the degree of nonlocality $w_0/\mu=100$.}
\end{figure}

\begin{figure}
\centering
\includegraphics[totalheight=3in]{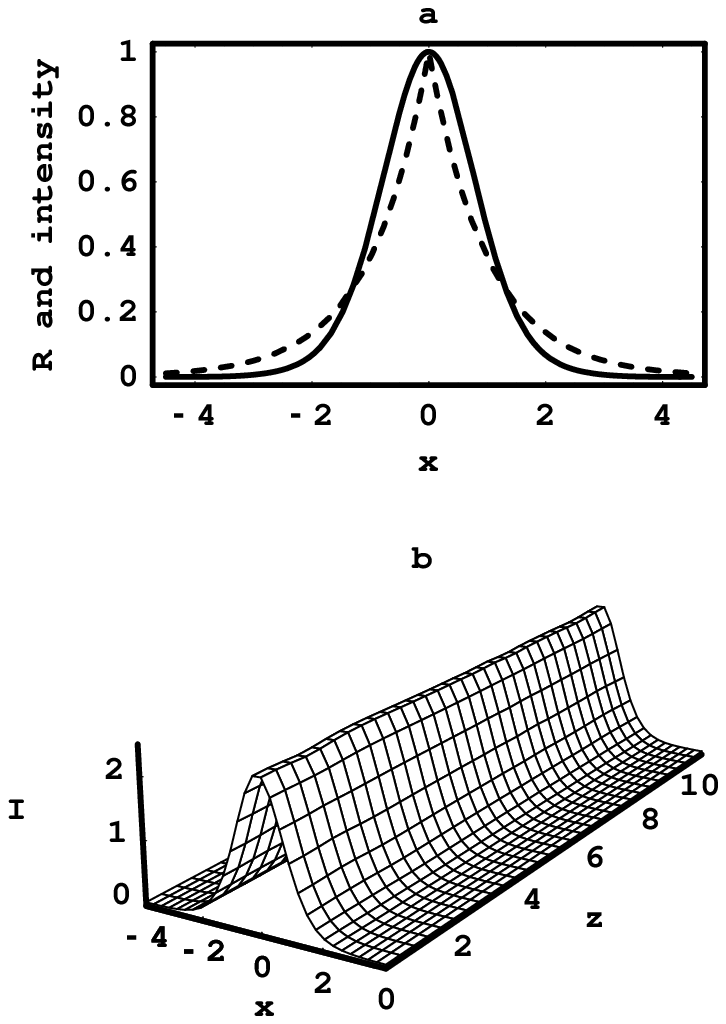}
\caption{\label{fig7}(a) The comparison between $R(x)$ and
$|\psi_0(A,\alpha,\beta,x)|^2$. Dashing line: $R(x)/R(0)$; solid
line: $|\psi_0(A,\alpha,\beta,x)|^2/|\psi_0(A,\alpha,\beta,0)|^2$;
(b)The propagation of the light beam with an input intensity
profile described by $|\psi_0(A,\alpha,\beta,x)|^2$. Here $w_0=1,
\mu=1, A=2.206, \alpha=-0.126, \beta=0.0280$ and the degree of
nonlocality $w_0/\mu=1$.}
\end{figure}

\begin{figure}
\centering
\includegraphics[totalheight=3in]{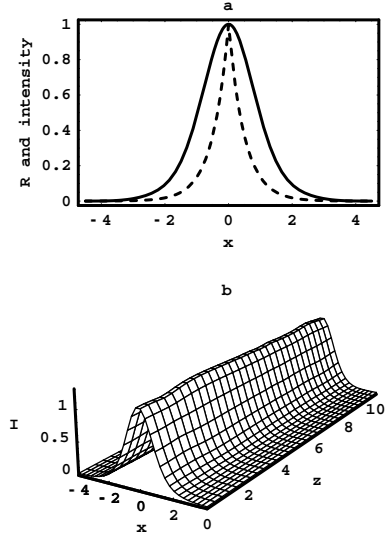}
\caption{\label{fig8}(a) The comparison between $R(x)$ and
$|\psi_0(A,\alpha,\beta,x)|^2$. Dashing line: $R(x)/R(0)$; solid
line: $|\psi_0(A,\alpha,\beta,x)|^2/|\psi_0(A,\alpha,\beta,0)|^2$;
(b) The propagation of the light beam with an input intensity
profile described by $|\psi_0(A,\alpha,\beta,x)|^2$. Here
$w_0=0.5, \mu=1, A=1.610, \alpha=-0.158, \beta=0.0408$ and the
degree of nonlocality $w_0/\mu=0.5$.}
\end{figure}

For this exponential-decay type nonlocal response and the
fundamental soliton state, $V(x)$ can be approximately given by
\begin{eqnarray}
&V(x)\approx-\int^{+\infty}_{-\infty}{{1}\over{2w_0}}\exp\left(-{{|x-\xi|}\over{w_0}}\right)|\psi_0(A,0,0,\xi)|^2d\xi~\nonumber\\
&={{A^2}\over{4w_0}}e^{{\mu^2
}\over{4w_0^2}}\{e^{-{{x}\over{w_0}}}[erf\left({{\mu}\over{2w_0}}-{{x}\over{\mu}}\right)-1]~~\nonumber\\
&+e^{{x}\over{w_0}}[erf\left({{\mu}\over{2w_0}}+{{x}\over{\mu}}\right)-1]\},
\end{eqnarray}
where
\begin{equation}
erf(x)={{2}\over{\sqrt{\pi}}}\int^x_0e^{-\xi^2}d\xi.
\end{equation}
Combining Eqs.~(\ref{parameters}), we get
\begin{subequations}
\begin{eqnarray}
&A^2\approx{{1/\mu}\over{{{\mu^2}\over{\sqrt{\pi}w_0^2}}+\exp({{\mu^2}\over{4w_0^2}}){{\mu^3}\over{2w_0^3}}[erf({{\mu}\over{2w_0}})-1]}},\\
&V_0\approx-{{A^2\exp({{\mu^2}\over{4w_0^2}})[1-erf({{\mu}\over{2w_0}})]}\over{2w_0}},\\
&\alpha\approx{{A^2\{\exp({{\mu^2}\over{4w_0^2}})[erf({{\mu}\over{2w_0}})-1]+{{2w_0}\over{\sqrt{\pi}\mu}}-{{4w_0^3}\over{\sqrt{\pi}\mu^3}}\}}\over{48w_0^5}},\\
&\beta\approx{{A^2\{\exp({{\mu^2}\over{4w_0^2}})[erf({{\mu}\over{2w_0}})-1]+{{2w_0}\over{\sqrt{\pi}\mu}}-{{4w_0^3}\over{\sqrt{\pi}\mu^3}}+{{24w_0^5}\over{\sqrt{\pi}\mu^5}}\}}\over{1440w_0^7}}.
\end{eqnarray}\label{parameters-exp}
\end{subequations}
In the strongly nonlocal limit the degree of nonlocality
$w_0/\mu\gg1$, we obtain
\begin{subequations}
\begin{eqnarray}
&A^2\approx{{\sqrt{\pi}w_0^2}\over{\mu^3}},\\
&V_0\approx-{{\sqrt{\pi}w_0}\over{2\mu^3}},\\
&\alpha\approx-{{1}\over{12\mu^6}},\\
&\beta\approx{{1}\over{60\mu^8}}.
\end{eqnarray}\label{p-exp-strongly}
\end{subequations}
It is worth to note that in the strongly nonlocal case the
parameters $\alpha$ and $\beta$ are free from the characteristic
nonlocal length $w_0$. Even when the characteristic nonlocal
length $w_0$ approaches infinity, the parameters $\alpha, \beta$
still rest on finite values and don't approach zero, and therefore
$\psi_0(A,\alpha,\beta,x)$ does't approach $\psi_0(A,0,0,x)$. That
greatly differs from the nonlocal case of a Gaussian function type
nonlocal response. As a result the Gaussian-function-like soliton
solution $\psi_0(A,0,0,x)$ can't describe the soliton state of the
NNLSE~(\ref{NNLSE}) exactly even in the strongly nonlocal case,
that is shown in Fig.~(\ref{fig6}). As shown in Fig.~(\ref{fig7}),
$\psi_0(A,\alpha,\beta,x)$ also can describe the soliton state of
the NNLSE~(\ref{NNLSE}) exactly when $w_0/\mu=1$. Even when
$w_0/\mu=0.5$, as indicated in Fig.~(\ref{fig8}),
$\psi_0(A,\alpha,\beta,x)$ can describe the soliton state of the
NNLSE~(\ref{NNLSE}) in high quality. As indicated by the values of
$\bar{\theta}$ in table~(\ref{table}), $\psi_0(A,\alpha,\beta,x)$
can describe the fundamental soliton states of NNLSE~(\ref{NNLSE})
exactly in the generally nonlocal case.

\begin{figure}
\centering
\includegraphics[totalheight=3in]{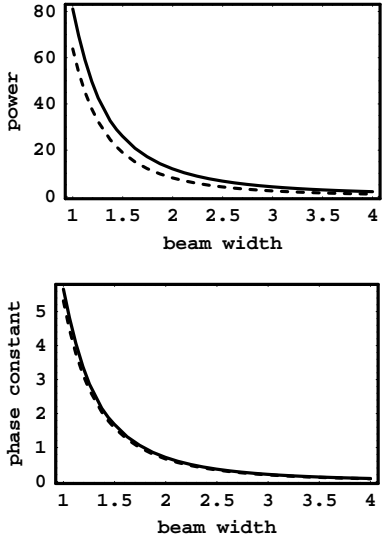}
\caption{\label{fig13}The dependence of the soliton's power $P$
and phase constant $\gamma$ on its beam width $\mu$. Dashing lines
are described by Eq.~(\ref{exp-P}) and Eq.~(\ref{exp-gamma})
respectively; Solid lines are directly calculated with parameters
$A,\alpha,\beta$ numerically found by the fixed-point method
presented in the section Appendix~(\ref{A}). Here $w_0=6$.}
\end{figure}

In the strongly nonlocal case the soliton's power and phase
constant are approximately given by
\begin{equation}
P\approx A^2\approx{{\sqrt{\pi}w_0^2}\over{\mu^3}},\label{exp-P}
\end{equation}
\begin{equation}
\gamma\approx-V_0\approx{{\sqrt{\pi}w_0}\over{2\mu^3}}\label{exp-gamma}
\end{equation}
respectively. For a given value of the characteristic nonlocal
length, the soliton's power and phase constant are both in inverse
proportion to the 3th power of its beam width in the strongly
nonlocal case that differs from the nonlocal case of a Gaussian
function type nonlocal response where the soliton's power and
phase constant are both in inverse proportion to the 4th power of
its beam width in the strongly nonlocal case. The dependence of
the soliton's power $P$ and phase constant $\gamma$ on its beam
width $\mu$ are shown in Fig.~(\ref{fig13}) for a given value of
characteristic nonlocal length. It is indicated that
Eq.~(\ref{exp-P}) and Eq.~(\ref{exp-gamma}) can describe these
dependence very well in the strongly nonlocal case.

\section{\label{third}the higher order generally nonlocal soliton solutions in the 2nd approximation}
\subsection{the nonlocal case of the Gaussian function type nonlocal response}
The second order soliton solution for the SNM~(\ref{SNM}) is given
by
\begin{equation}\label{solution-SNM2}
u_1(x,z)=A\left({{1}\over{\pi\nu^2}}\right)^{1/4}{{\sqrt{2}x}\over{\nu}}
e^{-{{x^2}\over{2\nu^2}}-i({{9}\over{4\nu^2}}-R_0A^2)z},
\end{equation}
where
\begin{equation}
{{1}\over{\nu^4}}=-R^{\prime\prime}_0A^2.
\end{equation}
The power and the beam width of $u_1(x,z)$ are given by $A^2$ and
$\sqrt{3}\nu$ respectively. For the Gaussian function type
nonlocal response, the second order soliton solution for the
SNM~(\ref{SNM}) is given by
\begin{equation}
u_1(x,z)=({{\sqrt{\pi}w_0^3}\over{2\nu^4}})^{{{1}\over{2}}}({{1}\over{\pi\nu^2}})^{{{1}\over{4}}}{{\sqrt{2}x}\over{\nu}}
e^{-{{x^2}\over{2\nu^2}}-i({{9}\over{4\nu^2}}-{{w_0^2}\over{2\nu^4}})z}.
\end{equation}
This soliton solution can describe the second order soliton state
of the NNLSE~(\ref{NNLSE}) exactly in the strongly nonlocal case
when $w_0/(\sqrt{3}\nu)>10$ but can't describe it exactly in the
generally nonlocal case when $w_0/(\sqrt{3}\nu)\sim2$.

\begin{figure}
\centering
\includegraphics[totalheight=1.2in]{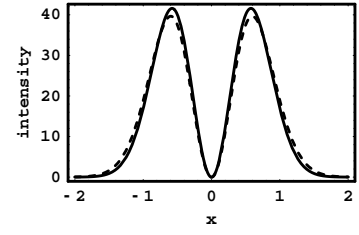}
\caption{\label{fig12}The comparison between
$|\psi_1(A,\alpha,\beta,x)|^2$ (dashing line) and
$|\psi_1(A,0,0,x)|^2$ (solid line). Here $w_0=1.5, \mu=1/\sqrt{3},
A=7.606, \alpha=-0.450, \beta=0.00530$ and the degree of
nonlocality $w_0/(\sqrt{3}\mu)=1.5$.}
\end{figure}

\begin{figure}
\centering
\includegraphics[totalheight=3in]{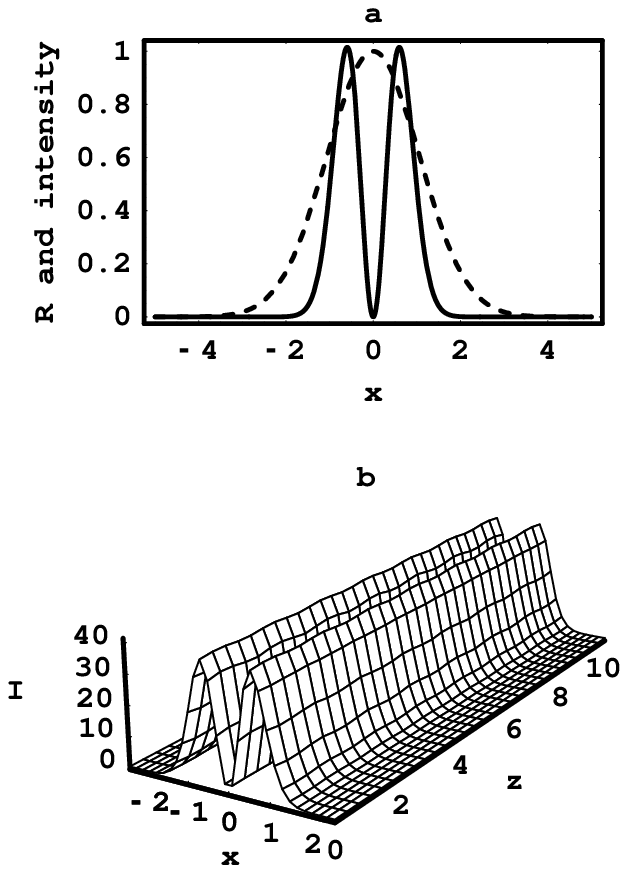}
\caption{\label{fig10}(a) The comparison between $R(x)$ and
$|\psi_1(A,\alpha,\beta,x)|^2$. Dashing line: $R(x)/R(0)$; solid
line:
$|\psi_1(A,\alpha,\beta,x)|^2/|\psi_1(A,\alpha,\beta,0.65)|^2$;
(b)The propagation of the light beam with an input intensity
profile described by $|\psi_1(A,\alpha,\beta,x)|^2$. Here
$w_0=1.5, \mu=1/\sqrt{3}, A=7.606, \alpha=-0.450, \beta=0.00530$
and the degree of nonlocality $w_0/(\sqrt{3}\mu)=1.5$.}
\end{figure}

\begin{figure}
\centering
\includegraphics[totalheight=3in]{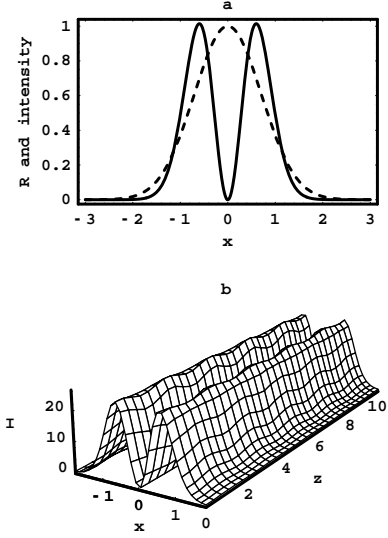}
\caption{\label{fig11}(a) The comparison between $R(x)$ and
$|\psi_1(A,\alpha,\beta,x)|^2$. Dashing line: $R(x)/R(0)$; solid
line:
$|\psi_1(A,\alpha,\beta,x)|^2/|\psi_1(A,\alpha,\beta,0.65)|^2$;
(b)The propagation of the light beam with an input intensity
profile described by $|\psi_1(A,\alpha,\beta,x)|^2$. Here $w_0=1,
\mu=1/\sqrt{3}, A=6.133, \alpha=-0.492, \beta=0.0145$ and the
degree of nonlocality $w_0/(\sqrt{3}\mu)=1$.}
\end{figure}

The second order generally nonlocal soliton solution in the 2nd
approximation is given by
\begin{eqnarray}\label{psi1}
&\psi_1(A,\alpha,\beta,x)\approx A({{1}\over{\pi\mu^2}})^{1/4}\exp(-{{x^2}\over{2\mu^2}}){{\sqrt{2}x}\over{\mu}}~~~~~~\nonumber\\
&\times[1+\alpha({{45\mu^6}\over{16}}-{{5\mu^4}\over{4}}x^2-{{\mu^2}\over{4}}x^4)~~~~~~~~~~~~~~~~~~~\nonumber\\
&+\alpha^2(-{{8375\mu^{12}}\over{512}}+{{215\mu^{10}}\over{64}}x^2+{{73\mu^8}\over{64}}x^4
+{{19\mu^6}\over{48}}x^6+{{\mu^4}\over{32}}x^8)\nonumber\\
&+\beta({{385\mu^8}\over{32}}-{{35\mu^6}\over{8}}x^2-{{7\mu^4}\over{8}}x^4-{{\mu^2}\over{6}}x^6)],~~~~~~~~~~~
\end{eqnarray}
and
\begin{equation}
\varepsilon_1\approx{{3}\over{2\mu^2}}+{{15\alpha\mu^4}\over{4}}-{{165\alpha^2\mu^{10}}\over{8}}+{{105\beta\mu^6}\over{8}}.\label{E1}
\end{equation}
For the Gaussian function type nonlocal response
function~(\ref{Gauss}), as shown in Fig.~(\ref{fig12}), the
difference between $|\psi_1(A,\alpha,\beta,x)|^2$ and
$|\psi_1(A,0,0,x)|^2$ is small in the generally nonlocal case. As
an Hermite-Gaussian function, the power and the beam width of
$\psi_1(A,0,0,x)$ are given by $A^2$ and $\sqrt{3}\mu$
respectively. So the power and the beam width of
$\psi_1(A,\alpha,\beta,x)$ are also approximatively given by $A^2$
and $\sqrt{3}\mu$ respectively. We can approximatively determine
the degree of nonlocality by $w_0/(\sqrt{3}\mu)$ and
approximatively abtain
\begin{eqnarray}
&V(x)\approx-\int^{+\infty}_{-\infty}{{1}\over{w_0\sqrt{\pi}}}\exp\left[-{{(x-\xi)^2}\over{w_0^2}}\right]\psi_1^2(A,0,0,\xi)d\xi\nonumber\\
&=-{{A^2}\over{\sqrt{\pi(\mu^2+w_0^2)}}}e^{-{{
x^2}\over{\mu^2+w_0^2}}}{{2x^2/\mu^2+w_0^2/\mu^2+w_0^4/\mu^4}\over{(1+w_0^2/\mu^2)^2}},\label{V1}
\end{eqnarray}
and
\begin{subequations}
\begin{eqnarray}
&A^2\approx{{\sqrt{\pi}(1+w_0^2/\mu^2)^{5/2}}\over{2\mu(w_0^2/\mu^2-2)}},\\
&V_0\approx-{{w_0^2(1+w_0^2/\mu^2)}\over{2\mu^4(w_0^2/\mu^2-2)}},\\
&\alpha\approx-{{(w_0^2/\mu^2-4)}\over{4\mu^6(1+w_0^2/\mu^2)(w_0^2/\mu^2-2)}},\\
&\beta\approx{{(w_0^2/\mu^2-6)}\over{12\mu^8(1+w_0^2/\mu^2)^2(w_0^2/\mu^2-2)}}.
\end{eqnarray}
\end{subequations}
In the strongly nonlocal limit the degree of nonlocality
$w_0/(\sqrt{3}\mu)\gg1$, we have
\begin{subequations}
\begin{eqnarray}
&A^2\approx{{\sqrt{\pi}w_0^3}\over{2\mu^4}},\\
&V_0\approx-{{w_0^2}\over{2\mu^4}},\\
&\alpha\approx-{{1}\over{4\mu^4w_0^2}},\\
&\beta\approx{{1}\over{12\mu^4w_0^4}}.
\end{eqnarray}
\end{subequations}
As the degree of nonlocality $w_0/(\sqrt{3}\mu)$ approaches
infinity, the parameters $\alpha$ and $\beta$ approach zero, and
$\psi_1(A,\alpha,\beta,x)$ approaches $\psi_1(A,0,0,x)$. Therefore
in the strongly nonlocal case an Hermite-Gaussian-function-like
second order soliton solution is obtained, and the power and the
phase constant of $\psi_1(A,\alpha,\beta,x)$ are both in inverse
proportion to the 4th power of its beam width. As indicated in
Fig.~(\ref{fig10}) and Fig.~(\ref{fig11}), the second order
soliton solution in the 2nd perturbation
$\psi_1(A,\alpha,\beta,x)$ can describe the second order soliton
state of the NNLSE~(\ref{NNLSE}) exactly when
$w_0/(\sqrt{3}\mu)=1.5$ and describe it in high qulity when
$w_0/(\sqrt{3}\mu)=1$. As shown by the values of $\bar{\theta}$ in
table~(\ref{table}),$\psi_1(A,\alpha,\beta,x)$ can exactly
describe the second order nonlocal soltion state in the generally
nonlocal cases.

Finally since the all eigenfunctions of the harmonic oscillator
can be found systematically\cite{24}, it is possible that in
analogy to a perturbed harmonic oscillator we can also
approximately calculate the third order soliton solution or the
fourth order soliton solution and so on in the generally nonlocal
case for the Gaussian function type nonlocal response.

\subsection{the nonlocal case of the exponential-decay type nonlocal response}
\begin{figure}
\centering
\includegraphics[totalheight=3in]{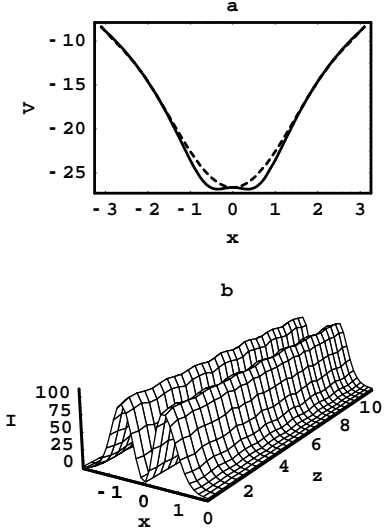}
\caption{\label{fig14}(a) The comparison between
$V(A,\alpha,\beta,x)$ (solid line) and $\tilde{V}(x)$ (dashing
line); (b)The propagation of the light beam with an input
intensity profile described by $|\psi_1(A,\alpha,\beta,x)|^2$.
Here $w_0=2, \mu=1/\sqrt{3}, A=12.162, \alpha=-0.442,
\beta=0.0179$ and the degree of nonlocality
$w_0/(\sqrt{3}\mu)=2$.}
\end{figure}

\begin{figure}
\centering
\includegraphics[totalheight=3in]{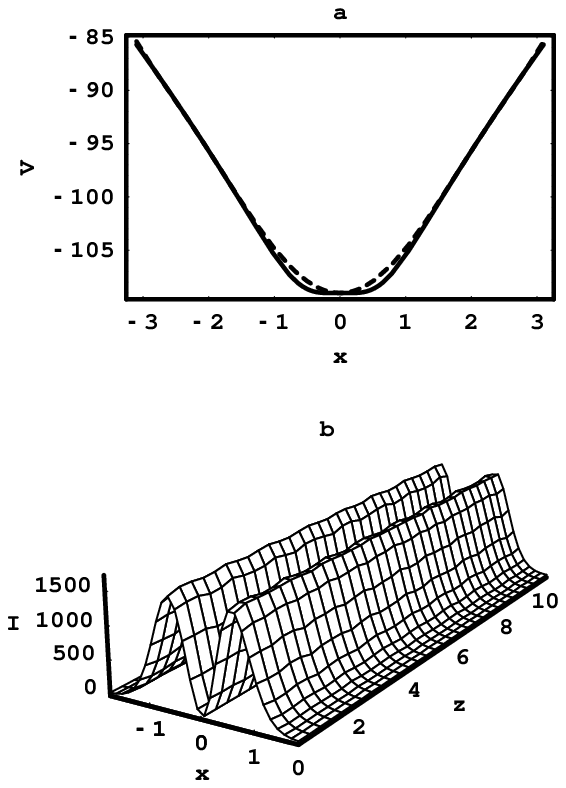}
\caption{\label{fig15}(a) The comparison between
$V(A,\alpha,\beta,x)$ (solid line) and $\tilde{V}(x)$ (dashing
line); (b)The propagation of the light beam with an input
intensity profile described by $|\psi_1(A,\alpha,\beta,x)|^2$.
Here $w_0=10, \mu=1/\sqrt{3}, A=48.257, \alpha=-0.348,
\beta=0.0141$ and the degree of nonlocality
$w_0/(\sqrt{3}\mu)=10$.}
\end{figure}

\begin{figure}
\centering
\includegraphics[totalheight=3in]{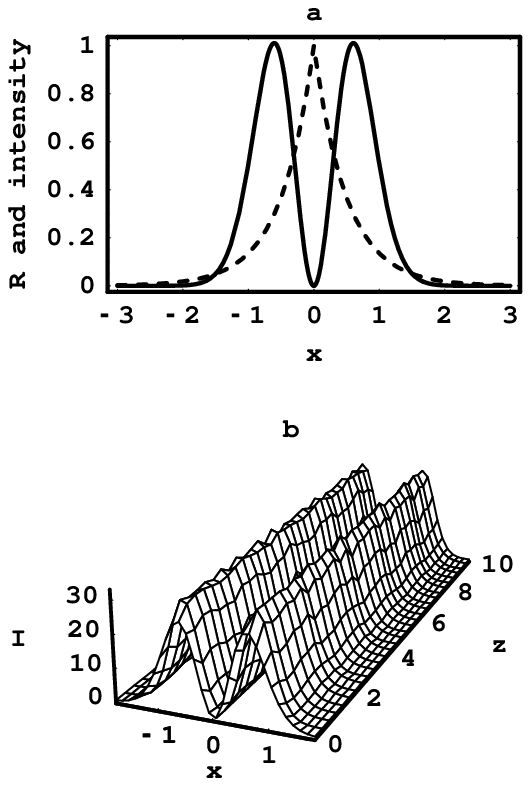}
\caption{\label{fig16}(a) The comparison between $R(x)$ and
$|\psi_1(A,\alpha,\beta,x)|^2$. Dashing line: $R(x)/R(0)$; solid
line:
$|\psi_1(A,\alpha,\beta,x)|^2/|\psi_1(A,\alpha,\beta,0.65)|^2$;
(b)The propagation of the light beam with an input intensity
profile described by $|\psi_1(A,\alpha,\beta,x)|^2$. Here
$w_0=0.5, \mu=1/\sqrt{3}, A=6.303, \alpha=-0.609, \beta=0.0270$
and the degree of nonlocality $w_0/(\sqrt{3}\mu)=0.5$.}
\end{figure}

As have been indicated, to $\psi_0(A,\alpha,\beta,x)$ and
$\psi_1(A,\alpha,\beta,x)$ in the nonlocal case of the Gaussian
function type nonlocal response and to $\psi_0(A,\alpha,\beta,x)$
in the nonlocal case of the exponential-decay type nonlocal
response, we have $V^{(2)}(A,\alpha,\beta,0)>0$ for generally
nonlocal cases. If we define
\begin{equation}
\tilde{V}(x)=V_0+{{1}\over{2\mu^4}}x^2+\alpha x^4+\beta x^6,
\end{equation}
we will get $V(A,\alpha,\beta,x)\approx \tilde{V}(x)$. But as
shown in Figs.~(\ref{fig14}a) and (\ref{fig15}a), to
$\psi_1(A,\alpha,\beta,x)$ in the nonlocal case of the
exponential-decay type nonlocal response, we have
$V^{(2)}(A,\alpha,\beta,0)<0$. In such a case we can't define
$1/\mu^4=V^{(2)}(A,\alpha,\beta,0)$ and can't define the
parameters $\mu,\alpha,\beta$ as those in Eqs.~(\ref{p-function}).
However as shown in Figs.~(\ref{fig14}a) and (\ref{fig15}a), we
still can find suitable values of $A,\alpha,\beta$ for a fixed
value of $\mu$ to make $V(A,\alpha,\beta,x)\approx \tilde{V}(x)$.
These suitable values of $A,\alpha,\beta$ can be calculated by
solving the following coupling equations
\begin{subequations}
\begin{eqnarray}
&V(A,\alpha,\beta,x_0)=V_0+{{1}\over{2\mu^4}}x_0^2+\alpha x_0^4+\beta x_0^6,\\
&V^{\prime}(A,\alpha,\beta,x_0)={{1}\over{\mu^4}}x_0+4\alpha x_0^3+6\beta x_0^5,\\
&V^{\prime\prime}(A,\alpha,\beta,x_0)={{1}\over{\mu^4}}+12\alpha
x_0^2+30\beta x_0^4,
\end{eqnarray}\label{parameters2}
\end{subequations}
where $x_0\neq0$ and $V_0=V(A,\alpha,\beta,0)$. In the section
Appendix~(\ref{B}) we present a fixed-point method to calculate
these parameters $A,\alpha,\beta$ with Eqs.~(\ref{parameters2}).
In Figs.~(\ref{fig14}b),(\ref{fig15}b) and (\ref{fig16}b) we show
the propagation of lights with input intensity profiles described
by $|\psi_1(A,\alpha,\beta,x)|^2$. Even when the
$w_0/(\sqrt{3}\mu)=0.5$, there still exists a second order
nonlocal soliton. As shown by the values of $\bar{\theta}$ in
table~(\ref{table}), $\psi_1(A,\alpha,\beta,x)$ can describe the
generally nonlocal soliton state in high quality. Since the
difference between $\psi_1(A,\alpha,\beta,x)$ and
$\psi_1(A,0,0,x)$ is small, we can approximately get
\begin{eqnarray}
&V(x)\approx-\int^{+\infty}_{-\infty}{{1}\over{2w_0}}\exp\left(-{{|x-\xi|}\over{w_0}}\right)|\psi_1(A,0,0,\xi)|^2d\xi~\nonumber\\
&={{A^2}\over{8w_0}}({{\mu^2}\over{w_0^2}}+2)e^{{\mu^2
}\over{4w_0^2}}\{e^{-{{x}\over{w_0}}}[erf({{\mu}\over{2w_0}}-{{x}\over{\mu}})-1]~~\nonumber\\
&+e^{{x}\over{w_0}}[erf({{\mu}\over{2w_0}}+{{x}\over{\mu}})-1]\}+{{A^2\mu}\over{2\sqrt{\pi}w_0^2}}e^{-{{x^2}\over{\mu^2}}}.\label{V2}
\end{eqnarray}
By defining
\begin{equation}
U(x)=V(x)/A^2\label{U}
\end{equation}
and combining with Eqs.~(\ref{parameters2}), we obtain
\begin{equation}
A\approx\sqrt{{{4x_0^2/\mu^4}\over{24[U(x_0)-U(0)]-9U^{\prime}(x_0)x_0+U^{\prime\prime}(x_0)x_0^2}}}.\label{A2}
\end{equation}
For example, when $w_0=10, \mu=1/\sqrt{3}, x_0=2$, from
Eqs.~(\ref{V2}), (\ref{U}) and (\ref{A2}) we get $A\approx47.323$
that is close to the numerically calculated value $A=48.257$.

While $V(A,\alpha,\beta,x)\approx\tilde{V}(x)$, as shown in
Figs.~(\ref{fig14}a) and (\ref{fig15}a) there still exists
difference $H(x)=V(A,\alpha,\beta,x)-\tilde{V}(x)$. To achieve
higher accuracy we should take $H(x)$ into account and set
$\tilde{V}(x)=V_0+x^2/(2\mu^2)+\alpha x^4+\beta x^6+H(x)$. Viewing
$H(x)$ as perturbation we will obtain another higher accurate
second order soliton solution. However the form of $H(x)$ is
rather complex and we will leave it in future further work and
don't intent to deal with the effect of $H(x)$ in this paper.

\begin{figure}
\centering
\includegraphics[totalheight=3in]{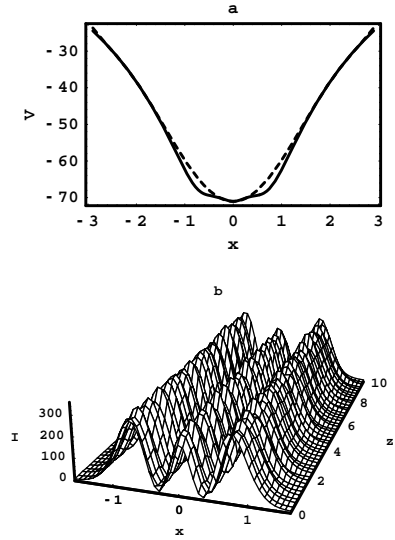}
\caption{\label{fig17}(a) The comparison between
$V(A,\alpha,\beta,x)$ (solid line) and $\tilde{V}(x)$ (dashing
line); (b)The propagation of the light beam with an input
intensity profile described by $|\psi_2(A,\alpha,\beta,x)|^2$.
Here $w_0=2, \mu=1/\sqrt{5}, A=19.694, \alpha=-1.327,
\beta=0.0606$ and the degree of nonlocality
$w_0/(\sqrt{5}\mu)=2$.}
\end{figure}

Now let us consider the third order nonlocal soliton. The third
order generally nonlocal soliton solution in the 2nd approximation
is given by
\begin{eqnarray}\label{psi2}
&\psi_2(A,\alpha,\beta,x)\approx A({{1}\over{\pi\mu^2}})^{1/4}\exp(-{{x^2}\over{2\mu^2}}){{1}\over{\sqrt{2}}}[-1+{{2x^2}\over{\mu^2}}\nonumber\\
&+\alpha(-{{45\mu^6}\over{16}}+{{123\mu^4}\over{8}}x^2-{{13\mu^2}\over{4}}x^4-{{1}\over{2}}x^6)+\alpha^2({{11927\mu^{12}}\over{512}}\nonumber\\
&-{{24587\mu^{10}}\over{256}}x^2+{{41\mu^8}\over{64}}x^4+{{193\mu^6}\over{96}}x^6+{{97\mu^4}\over{96}}x^8+{{\mu^2}\over{16}}x^{10})\nonumber\\
&+\beta(-{{655\mu^8}\over{32}}+{{1405\mu^6}\over{16}}x^2-{{125\mu^4}\over{8}}x^4-{{25\mu^2}\over{12}}x^6-{{1}\over{3}}x^8)].
\end{eqnarray}
As shown in Fig.~(\ref{fig17}) and
table~(\ref{table}),$\psi_1(A,\alpha,\beta,x)$ can describe the
third order generally nonlocal soliton only qualitatively. To
obtain a higher accurate third order soliton solution we should
take all perturbation into account or develop another new method.

\section{\label{conclusion}Conclusion}
In analogy to a perturbed harmonic oscillator, we calculate the
fundamental and some other higher order soliton solutions in the
2nd approximation in the generally nonlocal case. Numerical
simulations confirm that the soliton solutions in the 2nd
perturbation can describe the fundamental and second order soliton
states of the NNLSE~(\ref{NNLSE}) in high quality. For the
nonlocal case of the exponential-decay type nonlocal response, the
Gaussian-function-like soliton solution can't describe the
fundamental soliton state of the NNLSE~(\ref{NNLSE}) exactly even
in the strongly nonlocal case, that greatly differs from the
nonlocal case of the Gaussian function type nonlocal response. The
functional dependence of the nonlocal soliton's power and phase
constant on its beam width are greatly different from that of the
local soliton. In the strongly nonlocal case, the soltion's power
and phase constant are both in inverse proportion to the 4th power
of its beam width for the nonlocal case of the Gaussian function
type nonlocal response, and are both in inverse proportion to the
3th power of its beam width for the nonlocal case of the
exponential-decay type nonlocal response.

\begin{acknowledgments}
This research was supported by the National Natural Science
Foundation of China (Grant No. 10474023) and the Natural Science
Foundation of Guangdong Province, China(Grant No. 04105804).
\end{acknowledgments}

\appendix
\section{\label{A}how to calculate the parameters $A, \alpha, \beta$ with equations.~(\ref{p-function})}
In principle, $V(A,\alpha,\beta,x)$ and the parameters
$A,\alpha,\beta$ can be found by solving Eq.~(\ref{V-function})
and Eqs.~(\ref{p-function}) directly, but these tasks are
considerably involved. Here we present a fixed-point method to
calculate these parameters $A,\alpha,\beta$ for a fixed value of
$\mu$. Firstly corresponding to $V(A,\alpha,\beta,x)$ in
Eq.~(\ref{V-function}), we define
\begin{equation}
U(\alpha,\beta,x)={{V(A,\alpha,\beta,x)}\over{A^2}}.
\end{equation}
For an arbitrary pair of initial values of $\alpha_0,\beta_0$ with
suitable order of the magnitude, we can calculate
$U(\alpha_0,\beta_0,x)$. Let
\begin{eqnarray}
&A_1=\sqrt{{{1}\over{\mu^4U^{(2)}(\alpha_0,\beta_0,0)}}},\\
&\alpha_1=A_1^2 U^{(4)}(\alpha_0,\beta_0,0)/4!,\\
&\beta_1=A_1^2 U^{(6)}(\alpha_0,\beta_0,0)/6!.
\end{eqnarray}
For such a pair of values of $\alpha_1, \beta_1$, we can find
another $U(\alpha_1,\beta_1,x)$. Again we obtain another set of
values $\{A_2,\alpha_2,\beta_2\}$. Repeating these steps of
calculations, we can obtain series sets of values $\{A_2,
\alpha_3, \beta_3\}$, $\{A_3, \alpha_4, \beta_4\}$, and so on. The
difference between $\{A_m,\alpha_m,\beta_m\}$ and
$\{A_{m+1},\alpha_{m+1},\beta_{m+1}\}$ will approaches zero as the
number of $m$ approaches infinity. To some accuracy, we can
calculate parameters $A,\alpha,\beta$ for a fixed value of $\mu$.
\section{\label{B}how to calculate the parameters $A, \alpha, \beta$ with equations.~(\ref{parameters2})}
For a fixed value of $\mu$ and one suitable point $x_0\neq0$ (in
this paper we set $x_0=2$), corresponding to $V(A,\alpha,\beta,x)$
in Eq.~(\ref{V-function}) we define
\begin{equation}
U(x)={{V(A,\alpha,\beta,x)}\over{A^2}}.
\end{equation}
For an arbitrary pair of initial values of $\alpha_0,\beta_0$ with
suitable order of the magnitude, we can calculate $U(x)$. Let
\begin{eqnarray}
&A_1=\sqrt{{{4x_0^2/\mu^4}\over{24[U(x_0)-U(0)]-9U^{\prime}(x_0)x_0+U^{\prime\prime}(x_0)x_0^2}}},\\
&\alpha_1=A_1^2{{7U^{\prime}(x_0)x_0-12[U(x_0)-U(0)]-U^{\prime\prime}(x_0)x_0^2}\over{4x_0^4}},\\
&\beta_1=A_1^2{{U^{\prime\prime}(x_0)x_0^2+8[U(x_0)-U(0)]-5U^{\prime}(x_0)x_0}\over{8x_0^6}}.
\end{eqnarray}
For such a pair of values of $\alpha_1, \beta_1$, we can find
another $U(x)$. Again we obtain another set of values
$\{A_2,\alpha_2,\beta_2\}$. Repeating these steps of calculations,
to some accuracy we can calculate parameters $A,\alpha,\beta$ for
a fixed value of $\mu$.

　　　　　　　　　　　　　　　　　　　　　　　　　
\end{document}